\documentclass[fleqn,usenatbib]{mnras}

\usepackage{newtxtext,newtxmath}
\usepackage[T1]{fontenc}
\usepackage{ae,aecompl}

\usepackage{graphicx}	
\usepackage{amsmath}	
\usepackage{amssymb}	

\usepackage{booktabs}
\usepackage{threeparttable}
\usepackage[caption=false]{subfig}

\title[Pulse profile variation in PSR~B2217+47]{Low-frequency pulse profile variation in PSR~B2217+47: evidence for echoes from the interstellar medium}

\author[D.~Michilli et al.]{D.~Michilli,$^{1,2}$\thanks{E-mail: danielemichilli@gmail.com}
J.~W.~T.~Hessels,$^{1,2}$
J.~Y.~Donner,$^{3}$
J.-M.~Grie{\ss}meier,$^{4,5}$
M.~Serylak,$^{6,7}$\newauthor
B.~Shaw,$^{8}$
B.~W.~Stappers,$^{8}$
J.~P.~W.~Verbiest,$^{3,9}$
A.~T.~Deller,$^{10}$
L.~N.~Driessen,$^{1,8}$\newauthor
D.~R.~Stinebring,$^{11}$
L.~Bondonneau,$^{4}$
M.~Geyer,$^{12}$
M.~Hoeft,$^{13}$
A.~Karastergiou,$^{12,6,14}$\newauthor
M.~Kramer,$^{9,8}$
S.~Os{\l}owski,$^{10}$
M.~Pilia,$^{15,2}$
S.~Sanidas,$^{1}$
P.~Weltevrede$^{8}$
\\
$^{1}$Anton Pannekoek Institute for Astronomy, University of Amsterdam, Science Park 904, 1098 XH Amsterdam, The Netherlands\\
$^{2}$ASTRON, the Netherlands Institute for Radio Astronomy, Postbus 2, 7990 AA, Dwingeloo, The Netherlands\\
$^{3}$Fakult\"{a}t f\"{u}r Physik, Universit\"{a}t Bielefeld, Postfach 100131, 33501 Bielefeld, Germany\\
$^{4}$LPC2E - Universit\'{e} d'Orl\'{e}ans / CNRS, 45071 Orl\'{e}ans cedex 2, France\\
$^{5}$Station de Radioastronomie de Nan\c{c}ay, Observatoire de Paris, PSL Research University, CNRS, Univ. Orl\'{e}ans, OSUC, 18330 Nan\c{c}ay, France\\
$^{6}$Department of Physics \& Astronomy, University of the Western Cape, Private Bag X17, Bellville 7535, South Africa\\
$^{7}$SKA South Africa, 3rd Floor, The Park, Park Road, Pinelands, 7405, South Africa\\
$^{8}$Jodrell Bank Centre for Astrophysics, School of Physics and Astronomy, The University of Manchester, Manchester M13 9PL, UK\\
$^{9}$Max-Planck-Institut f\"{u}r Radioastronomie, Auf dem H\"{u}gel 69, 53121 Bonn, Germany\\
$^{10}$Centre for Astrophysics and Supercomputing, Swinburne University of Technology, P.O. Box 218, Hawthorn, VIC 3122, Australia\\
$^{11}$Dept. of Physics and Astronomy, Oberlin College, 110 North Professor St., Oberlin, OH 44074, USA\\
$^{12}$Oxford Astrophysics, Denys Wilkinson Building, Keble Road, Oxford OX1 3RH, UK\\
$^{13}$Th\"{u}ringer Landessternwarte Tautenburg, Sternwarte 7, 07778 Tautenburg, Germany\\
$^{14}$Department of Physics and Electronics, Rhodes University, PO Box 94, Grahamstown 6140, South Africa\\
$^{15}$INAF - Osservatorio Astronomico di Cagliari, via della Scienza 5, 09047 Selargius (Cagliari), Italy
}

\date{Accepted XXX. Received YYY; in original form ZZZ}
\pubyear{2017}

\begin{document}
\label{firstpage}
\pagerange{\pageref{firstpage}--\pageref{lastpage}}
\maketitle

\begin{abstract}
We have observed a complex and continuous change in the integrated pulse profile of PSR~B2217+47, manifested as additional components trailing the main peak.  
These transient components are detected over 6 years at $150$\,MHz using the LOw Frequency ARray (LOFAR), but they are not seen in contemporaneous Lovell observations at $1.5$\,GHz.
We argue that propagation effects in the ionized interstellar medium (IISM) are the most likely cause.
The putative structures in the IISM causing the profile variation are roughly half-way between the pulsar and the Earth and have transverse radii $R \sim 30$\,AU.
We consider different models for the structures.
Under the assumption of spherical symmetry, their implied average electron density is $\overline{n}_e \sim 100$\,cm$^{-3}$.
Since PSR~B2217+47 is more than an order of magnitude brighter than the average pulsar population visible to LOFAR, similar profile variations would not have been identified in most pulsars, suggesting that subtle profile variations in low-frequency profiles might be more common than we have observed to date.  
Systematic studies of these variations at low frequencies can provide a new tool to investigate the proprieties of the IISM and the limits to the precision of pulsar timing. 
\end{abstract}


\section{Introduction}\label{sec:introduction}
When averaged over hundreds of rotational periods, pulsars typically show stable integrated pulse profiles over timescales of years to decades \citep{Hel75,Liu12}.  
The stability of the integrated pulsar emission is key to using pulsars as `astrophysical clocks' in timing experiments \citep{Man17}.
However, subtle, long-timescale variations sometimes exist,
and can be either intrinsic to the pulsar magnetosphere \citep{Hob10}
or due to varying propagation effects as the signal travels through the ionised interstellar medium \citep[IISM,][]{Kei13}.  
Studying pulse profile changes in radio
pulsars is thus motivated by both understanding the underlying
physical mechanisms responsible for the observed changes and by
improving the precision of pulsar timing experiments.

Only a small fraction of isolated pulsars have been observed to manifest a continuous pulse profile evolution.
\citet{Sta00} attributed the quasi-periodic profile variation of PSR~B1828$-$11 to free precession of the neutron star. 
\citet{Lyn10} questioned this interpretation when they found quasi-periodic profile changes in six pulsars (including PSR~B1828$-$11), which are correlated with the spin-down rate and thus likely originate from processes intrinsic to the source.  
External factors have also been invoked to explain observed profile variations: e.g, \citet{Kar11} reported a pulse profile variation for PSR~J0738$-$4042 attributed to magnetospheric changes, which \citet{Bro14} connected to an interaction with an asteroid.  
A systematic search by \citet{Bro16} found seven examples of profile changes in a sample of 168 pulsars, a subset of which were correlated with $\dot{\nu}$ variations.
Os{\l}owski et al. (in prep.) observed a profile variation of PSR~B1508+55 presenting characteristics similar to the one reported here for PSR~B2217+47.
They attribute the variation to IISM propagation effects.


Pulsar signals propagate through the IISM before reaching the Earth,
and this imparts several features on the observed signal
\citep[e.g.][]{Ric90}.  Among these, dispersion is the
frequency-dependent light travel time due to free electrons along the
line of sight (LoS), where the integrated column density is quantified by
the dispersion measure \citep[DM,][]{Lor04}.
A second propagation effect is scintillation and it is due to inhomogeneities in the IISM electron density \citep{Ric90}.
A peculiar manifestation of propagation effects is an extreme scattering event \citep[ESE,][]{Fie87}.
During an ESE, a dense plasma structure of finite size crosses the LoS to a point source.  This can lead to different detectable variations in the signal \citep[e.g. in the source flux,][]{Fie87,Cog93}. 
Thus far, variations in a pulsar's average profile due to dense structures crossing the LoS have only been reported in the form of echoes produced by the Crab nebula surrounding PSR~B0531+21 \citep{Bac00,Lyn01}.
Another manifestation of propagation effects in pulsar observations are scintillation arcs.
These are parabolic arcs visible in some pulsars' secondary spectra, i.e. the two-dimensional Fourier transform of the signal as a function of frequency and time \citep{Sti01}.
They are thought to be produced by scintillation caused by thin screens along the LoS.


In recent years, a new generation of radio telescopes such as the LOw Frequency ARray \citep[LOFAR,][]{Haa13}, the Murchison Widefield Array \citep[MWA,][]{Tin13} and the Long Wavelength Array \citep[LWA,][]{Tay12} have renewed interest in pulsar studies at frequencies below $300$\,MHz, including investigations of pulse profiles.
Low-frequency studies potentially allow for more sensitive analyses of profile variability, both if the cause is intrinsic to the pulsar emission or due to propagation effects.
In fact, the radius-to-frequency-mapping model \citep{Rud75,Cor75} states that lower-frequency radio waves are emitted at higher altitudes above the neutron star's magnetic poles, which implies a larger cone of emission at lower frequencies and hence amplified angular-dependent variations in pulsar beams.
Likewise, low-frequency radio waves are far more sensitive to propagation effects in the IISM.

Here we present a comprehensive study of profile variations in
PSR~B2217+47 (J2219+4754), a slow pulsar discovered by \citet{Tay69}.
With a mean flux density of $820\pm410$\,mJy at $150$\,MHz \citep{Bil16}, PSR~B2217+47 is one of the brightest pulsars in the low-frequency sky.
This allows even subtle profile changes to be detected.
The pulse profile of PSR~B2217+47 is typically single-peaked below
$300$\,MHz \citep[e.g.][]{Kuz98}.  However, we noticed that the
single-epoch LOFAR $150$\,MHz profile reported in \citet{Pil16} shows
a prominent trailing secondary
component\footnote{\url{www.epta.eu.org/epndb}}.  
Intriguingly, using the BSA telescope at the Pushchino Observatory at a central frequency of 102.5\,MHz, \citet{Sul94} previously reported variations in the pulse profile of PSR~B2217+47.
They detected a similar secondary component appearing and evolving between 1983~--~1984 and again between 1987~--~1992.  This feature was not detected at $325$\,MHz in
2006~--~2007 \citep{Mit11}, nor was it seen by \citet{Bas15}; the authors report the absence of the transient component in profiles at higher frequency but they do not give additional information on the observations.
\citet{Sul94} attributed the evolving component to pulsar free
precession.  As an alternative explanation, significant DM changes detected towards PSR~B2217+47
suggest a strongly inhomogeneous IISM along the pulsar's
LoS \citep{Ahu05}.  The DM of the source increased by $0.02$\,pc\,cm$^{-3}$ over $\sim400$ days in 2001~--~2002, decreasing
again in the following $\sim100$ days \citep{Ahu05}.  
This was the only systematic DM variation among a
sample of 12 pulsars reported by \citet{Ahu05}.  


We performed a dense, multi-frequency campaign, with LOFAR observations beginning in 2011.  
The observations used in this study are described in \textsection\ref{sec_observations} along with the processing methods employed.
In \textsection\ref{sec:results} we analyse the changes in the pulsar characteristics as a function of time.
Different scenarios for the origin of these variations and their implications are discussed in \textsection\ref{sec:discussion}.
An IISM interpretation is favoured by the observations and we develop the model in \textsection\ref{sec:IISM}.
We conclude the study by summarizing our findings in \textsection\ref{sec:conclusions}.

\section{Observations and data processing}\label{sec_observations}

\begin{table*}
\centering
\begin{threeparttable}
\caption{Summary of the observations used to form our data set. TOAs were calculated for all the observations except where indicated otherwise. Pulse profiles were obtained from all the observations after 2011 March.
}
\label{tab:observations}
\begin{tabular}{llllll}
\toprule
Telescope & Centre frequency & Bandwidth	& Typical integration	& N. obs. & Timespan \\
& (MHz) & (MHz) & time (min) & & \\
\midrule
LOFAR core	 & 122 to 151 & 2 to 92 & 5 & 17* & 2011 Mar -- 2012 Nov \\ 
LOFAR core	 & 149	& 78 & 10 & 40 & 2013 Dec -- 2017 Feb \\ 
\multicolumn{6}{c}{}\\
DE601     & 145 & 36 & 15 & 23 & 2013 Jun -- 2013 Jul \\
DE601     & 139 & 48 & 30 & 7 & 2013 May -- 2013 Aug \\
DE601     & 157 & 61 & 20 & 2* & 2013 Aug -- 2013 Sep \\
DE601     & 154 & 54 & 13 & 2* & 2013 Dec \\
DE601     & 149 & 78 & 90 & 18 & 2013 Aug -- 2015 Jun \\
\multicolumn{6}{c}{}\\
DE603     & 149 & 78 & 10 & 10 & 2014 Feb -- 2014 May \\
\multicolumn{6}{c}{}\\
DE605     & 149 & 78 & 70 & 9 & 2014 Dec -- 2015 Jan \\
DE605     & 158 & 64 & 100 & 22 & 2015 May -- 2015 Jul \\
\multicolumn{6}{c}{}\\
FR606     & 162 & 48 & 30 & 9 & 2014 Feb -- 2014 May \\
FR606     & 149 & 78 & 120 & 1* & 2014 May \\
\multicolumn{6}{c}{}\\
UK608     & 167 & 48 & 60 & 11* & 2013 Jun -- 2013 Nov \\
UK608     & 161 to 164 & 36 & 60 & 3* & 2013 Jul -- 2015 May \\
UK608     & 162 & 48 & 60 & 22 & 2013 Jun -- 2015 Oct \\
\multicolumn{6}{c}{}\\
Mark II & 1400 & 32 & 10 & 56 & 2002 Mar -- 2003 Aug \\
Lovell & 235 & 1, 4 & 20 & 3 & 1984 Sep -- 1987 Dec \\
Lovell & 325 & 8 & 10 & 2 & 1995 Feb -- 1996 Mar \\
Lovell & 410 & 1, 2, 4, 8 & 10 & 54 & 1984 Sep -- 1997 Sep \\
Lovell & 610 & 1, 4, 8 & 10 & 70 & 1984 Sep -- 2007 Oct \\
Lovell & 925 & 8 & 10 & 6 & 1989 Feb -- 1989 Aug \\
Lovell & 1400 & 32, 40, 96 & 10 & 410 & 1984 Aug -- 2009 Aug \\
Lovell & 1520 & 384 & 5 & 128 & 2009 Aug -- 2017 Feb \\
Lovell & 1625 & 32, 40 & 10 & 11 & 1989 Feb -- 1994 Nov \\
\multicolumn{6}{c}{}\\
Goldstone DSS 13, 14 & 2388 & 12 & >5 & 118 & 1969 Dec -- 1982 May \\
\bottomrule
\end{tabular}
\begin{tablenotes}
\item[*] TOAs were not calculated for these observations
\end{tablenotes}
\end{threeparttable}
\end{table*}

The analysis presented in this paper is based on the observations summarised in Table~\ref{tab:observations}.
Times of arrival (TOAs) between December 1969 and May 1982 were obtained by \citet{Dow83} using the the NASA Deep Space Network.
Jodrell Bank observations span $32$ years, primarily using the $76$-m Lovell telescope.
Occasional supplementary observations were made using the $38\times25$-m Mark II telescope.  
They are included in the Jodrell Bank data archive of pulsar observations \citep{Hob04}.
Detailed information on Lovell and Mark II observations and data analysis are provided by \citet{She96}, \citet{Gou98} and \citet{Hob04}.
LOFAR observations span $6$ years.
Pulsar observing with LOFAR is extensively described by \citet{Sta11}.
All the observations presented here have been acquired using the
high-band antennas (HBAs), which observe at a central frequency of
$\sim150$\,MHz.  
Low-band antenna (LBA) observations, which have a central frequency of $\sim 50$\,MHz, were not used because PSR~B2217+47 is heavily scattered at these frequencies \citep{Pil16} and this masks the subtle profile variations studied here.
The scattering itself might also be variable, but we were unable to investigate this due to the small number of available LBA observations and their limited sensitivity.
We used observations acquired with both the LOFAR core and international stations.
We used 5 international LOFAR stations in stand-alone (local recording) mode: DE601 in Effelsberg, DE603 in Tautenburg, DE605 in J{\"u}lich, FR606 in Nan{\c c}ay and UK608 in Chilbolton.
The observations were taken using different telescope configurations and different time and frequency resolutions, as summarised in Table~\ref{tab:observations}.

Jodrell Bank observations were processed following \citet{Gou98} for early observations and \citet{Hob04} for more recent observations.
A single pulse profile and one TOA were obtained for each observation.
We omitted RFI-contaminated profiles by visual inspection.
Only those pulse profiles obtained during the LOFAR campaign have been used in this study.
We processed the LOFAR observations using the \textsc{psrchive}\footnote{\url{psrchive.sourceforge.net}} software library \citep{Hot04,Str12}.
All the programs mentioned in the following paragraph are part of this package unless indicated otherwise.
As a first step, we mitigated radio frequency interference (RFI) present in the datasets using both \textsc{paz} and \textsc{clean.py} offered by the \textsc{coast guard} package\footnote{\url{github.com/plazar/coast\_guard}}\citep{Laz16}.
Subsequently, we folded each dataset using an initial timing solution
obtained by \citet{Hob04}.
Small DM variations on timescales of days can be non-negligible at these low frequencies. 
Therefore, we initially calculated the best DM value for each observation by maximising the integrated pulse profile's signal-to-noise ratio (S/N) using \textsc{pdmp}.  
One total-intensity profile was obtained from each LOFAR observation. 
For each LOFAR station, we used one template of the pulse profile for each group of observations at the same frequency.
The templates were obtained by fitting the main peak of a high S/N profile with a single von Mises function using \textsc{paas}.
A TOA for each profile was calculated using \textsc{pat}.
Highly precise DM values were calculated from observations performed with the core and German stations using frequency-resolved timing. 
The details of this analysis are described in our companion paper by Donner et al. (in prep.) and the resulting DM time series is replicated in Fig~\ref{fig:LOFAR_DM_flux}a. 
These DM values were applied in our work to correct for the time-variable dispersion both in our profile-shape investigations and in our long-term timing analysis.
TOAs were not calculated for early commissioning and problematic observations (e.g. affected by strong interference or by software failures).  
In order to study the pulsar's profile evolution, the baseline was subtracted from each profile and each profile was normalised to a peak amplitude of unity.
Pulse profiles were aligned by cross-correlating the main peak with one single-peaked template.

The 15-minute observation reported by \citet{Pil16} has been used extensively in this study.
It is the first high-quality LOFAR observation available, dating back to 2011 October 24, and (coincidentally) was recorded almost simultaneously with the one known rotational glitch of PSR~B2217+47 \citep{Esp11}.  
This observation has a central frequency of $143$\,MHz and a bandwidth of $47$\,MHz divided into $3840$ channels.
It shows the transient component well separated from the main peak.  
It is also one of the few observations for which we have single-pulse data.

\section{Data analysis and results}\label{sec:results}

\subsection{Timing analysis}

\begin{table}
\centering
\begin{threeparttable}
\caption{Ephemeris obtained for PSR~B2217+47.
Uncertainties in parentheses refer to the last quoted digits. 
}	
\label{tab:ephemeris}
\begin{tabular}{ll}
\toprule
Parameter & Value \\
\midrule
Right ascension RA (J2000) (h:m:s)                       & $22$:$19$:$48.128(4)$  \\
Declination DEC  (J2000) (\degr:\arcmin:\arcsec)             & $+47$:$54$:$53.82(4)$ \\
RA Proper Motion (mas\,yr$^{-1}$)         & $-12(3)$              \\
DEC Proper Motion (mas\,yr$^{-1}$)        & $-19(3)$              \\
Position Epoch (MJD)                     & $49195$                   \\
\multicolumn{2}{c}{}\\
Spin period $P$ (s)                               & $0.53846945053(2)$     \\
Spin period derivative $\dot{P} $($10^{-15}$)                   & $2.76516(6)$     \\
Spin frequency $\nu$ (Hz)                               & $1.85711556896(7)$     \\
Spin frequency derivative $\dot{\nu}$ ($10^{-15}$ s$^{-2}$)        & $-9.5367(2)$          \\
Period Epoch (MJD)                       & $49195$                   \\
\multicolumn{2}{c}{}\\
DM (pc\,cm$^{-3}$)                        & $43.517(3)$             \\
DM Epoch (MJD)                           & $49195$                   \\
\multicolumn{2}{c}{}\\
Glitch Epoch (MJD)                       & $55859.43$\tnote{a} \\
Glitch $\Delta\nu$ ($10^{-9}$ Hz)        & $2.15(2)$              \\
Glitch $\Delta\dot{\nu}$ ($10^{-18}$ s$^{-2}$) & $-5(1)$        \\
\multicolumn{2}{c}{}\\
Range of observations (MJD)		&$40585$ -- $57806$		\\
\multicolumn{2}{c}{}\\
Surface magnetic field B (G)				& $1.2 \times 10^{12}$				\\
Characteristic age $\tau$ (Myr)				& $3.1$				\\
Spin-down energy $\dot{\text{E}}$ (erg\,s$^{-1}$) & $7.0 \times 10^{32}$				\\
\bottomrule
\end{tabular} 
\begin{tablenotes}
\item[a] Value from \citet{Esp11}
\end{tablenotes}
\end{threeparttable}
\end{table}

\begin{figure*}
\centering
\includegraphics{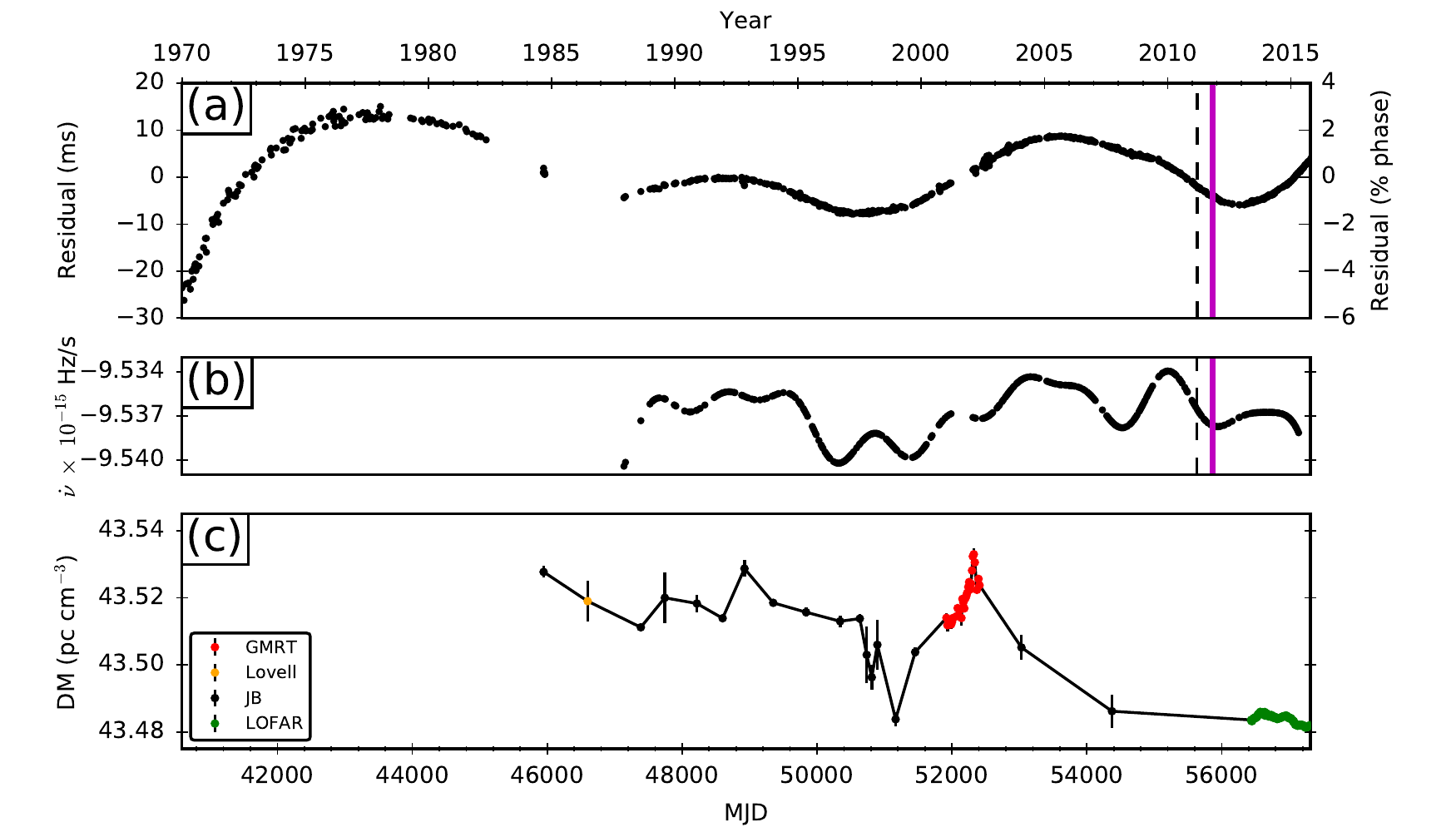}
\caption{Residuals of the fit for the timing model reported in Table~\ref{tab:ephemeris} (a) and spin-down variations (b).
For both panels, the vertical dashed line indicates the first LOFAR observation and the vertical magenta line the epoch of a rotational glitch.
Long-term DM variations for the source are shown in panel c. 
Giant Metrewave Radio Telescope (GMRT) values (red) are from \citet{Ahu05} and the orange point was obtained by \citet{Hob04} using the Lovell telescope.
Black and green points are unpublished values from Jodrell Bank (JB) and LOFAR (core and German stations), respectively.  
}
\label{fig:timing}
\end{figure*}

All available TOAs have been fitted using \textsc{tempo2}\footnote{\url{bitbucket.org/psrsoft/tempo2}} \citep{Hob06} to refine the timing model.
TOAs before Modified Julian Date (MJD)~45095 are from the NASA Deep Space Network and show an additional scatter in their residuals that is much larger than the quoted statistical uncertainties.  
To account for (the assumed) systematics in the data, we assigned these TOAs an equal uncertainty of 1\,ms estimated from the scatter of their values.
The strong timing noise complicated a global fit.  For this reason, we used the Cholesky method presented by \citet{Col11} to whiten the spectrum.
We included in our model the one known rotational glitch of the pulsar \citep{Esp11}.
The parameters obtained from this fit are presented in Table~\ref{tab:ephemeris} and the timing residuals are shown in Fig.~\ref{fig:timing}a.  
The residuals show a slow modulation over the
considered time span, compatible with the analysis presented by \citet{Cor80}, \cite{Hob10} and \citet{Sha13}. 

It was possible to verify the accuracy of our timing analysis using an independent measurement of the source position obtained from an imaging observation with the Very Long Baseline Array (VLBA) from the PSR$\pi$ pulsar astrometry campaign \citep{Del11}.
Assuming an error dominated by systematic uncertainties on the position of PSR~B2217+47 of 5\,mas \citep{Del16}, the obtained right ascension is 22h19m48.1070(5)s and the declination is +47\degr54\arcmin53.471(5)\arcsec on MJD~55569, consistent with the timing analysis after accounting for the measured proper motion.


In order to investigate a possible connection between the spin-down $(\dot{\nu})$ evolution of this pulsar and its pulse-shape variations, we used the method developed by \citet{Bro16} to compute $\dot{\nu}(t)$. 
This method uses Gaussian Process Regression (GPR) to model timing residuals and profile variations.
We found that the residuals of PSR~B2217+47 were best fit by a single squared exponential kernel \citep{Ras06} chosen due to its differentiable properties. To allow for uncertainty on the residuals, we used an additional white noise kernel. 
The fit is applied to the residuals by optimising the hyper-parameters $\theta ( \lambda, \sigma^2, \sigma_n^2)$ associated with the kernels, where $\lambda$ is the function smoothness, $\sigma^2$ is the average distance of the function from its mean value and $\sigma_n^2$ specifies the noise variance.
The optimised hyper-parameters were $\lambda = 1000$ days, $\sigma^2 = 1.4 \times 10^{-4}$ s and $\sigma_n^2 = 2.1 \times 10^{-7}$ s. 
The resulting spin-down evolution is shown in Fig.~\ref{fig:timing}b, where oscillations are visible on timescales of the order of 3 years. We find that the overall trend is approximately constant with a peak-to-peak fractional amplitude $\Delta \dot{\nu} / \dot{\nu} < 0.1\%$. The low quality of early TOAs did not permit to perform the analysis before 1988.

\subsection{DM and flux variations}\label{sec:results_DM}

\begin{figure}
\centering
\includegraphics{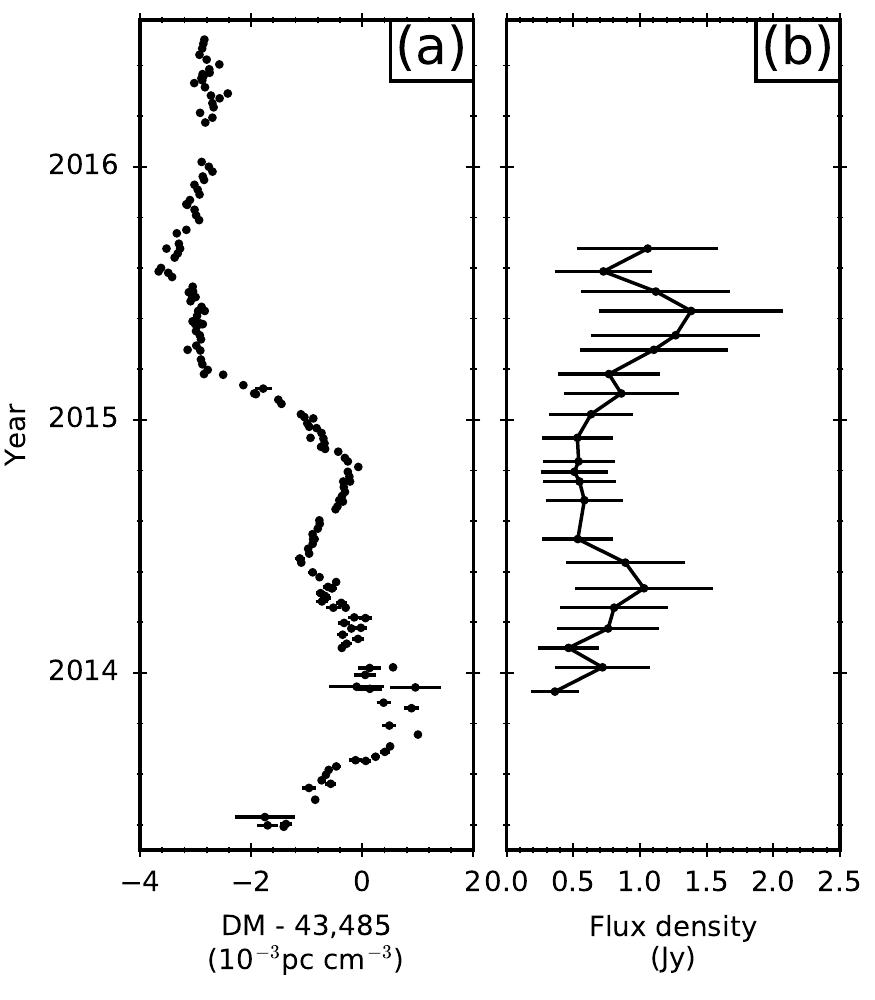}
\caption{DM (a) and mean flux density (b) of PSR~B2217+47 measured by LOFAR.
1-$\sigma$ error bars are often smaller than the points in panel (a), which is adapted from Donner et al. (in prep.).
}
\label{fig:LOFAR_DM_flux}
\end{figure}

In our companion paper (Donner et al., in prep.), we present dramatic variations in the DM towards this pulsar, as measured from the LOFAR data taken with the international stations in Germany (i.e. the the German Long Wavelength Consortium, GLOW).
In Fig.~\ref{fig:timing}c, these DM values are plotted together with archival Lovell data (see Table~\ref{tab:observations} for details), and other published values\footnote{\citet{Sto15} reported a significantly different DM value using LWA1.  
We attribute this offset to the strong scattering of the pulsar at $50$\,MHz.}. 
DM variations obtained from GLOW and LOFAR core observations are further highlighted in Fig.~\ref{fig:LOFAR_DM_flux}a.
The level of DM variations derived from LOFAR data are $\sim 5\times10^{-3}$\,pc\,cm$^{-3}$, roughly an order of magnitude lower than that reported by \citet{Ahu05}.
However, they are still significant compared to more typical DM variations observed along other lines of sight \citep[e.g.][and references therein]{Kei13}.

\begin{figure}
\centering
\includegraphics{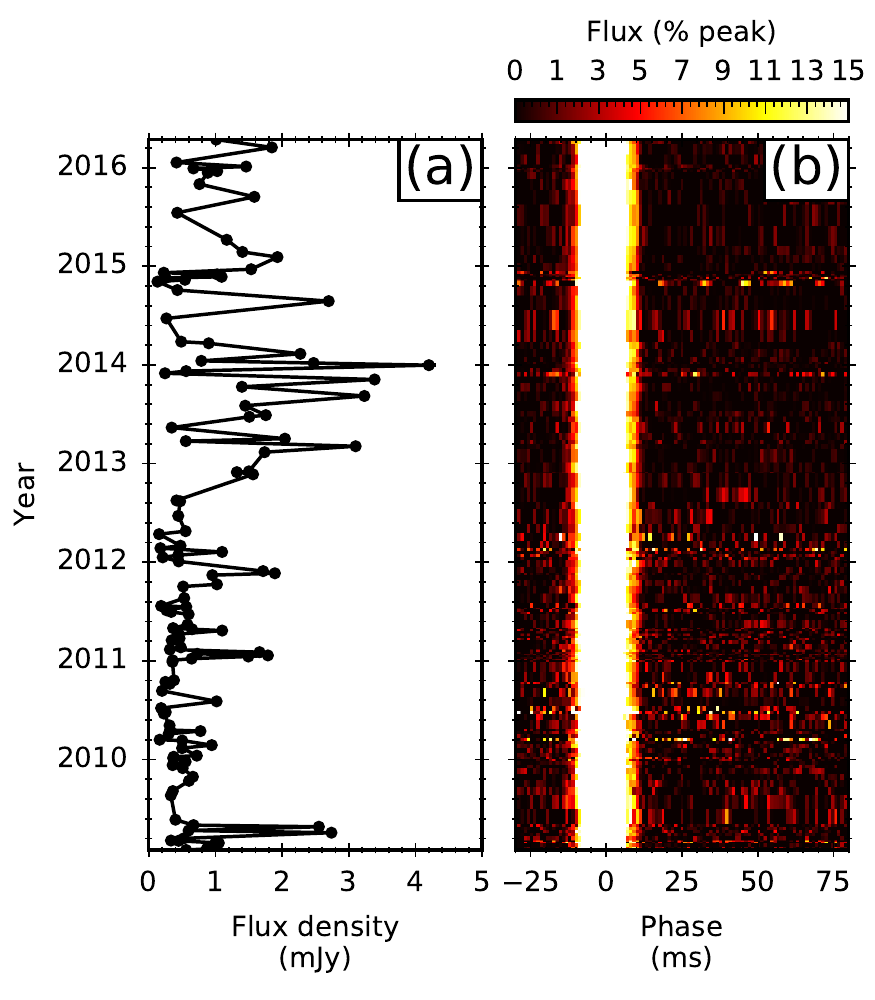}
\caption{Mean flux density (a) and pulse profiles (b) of PSR~B2217+47 measured at $1.5$\,GHz by the Lovell telescope. 
All the profiles are normalised to a peak flux of unity and aligned by cross-correlating with a standard template (and saturated to 15 per cent of the peak intensity).
1-$\sigma$ error bars are often smaller than the points in panel (a).
}
\label{fig:Lovell}
\end{figure}

A total of 23 observations obtained from the LOFAR core were
suitable for flux and polarisation calibration. 
We followed the procedure described in detail by \citet{Nou15} and \citet{Kon16}.  
The uncertainties of the resulting flux densities are conservatively estimated to be 50 per cent \citep{Bil16}. 
The resulting values are plotted in Fig.~\ref{fig:LOFAR_DM_flux}b.
Flux densities of recent Lovell observations are reported in Fig.~\ref{fig:Lovell}a. 
The scatter of data points is due to scintillation.

\begin{figure*}
\centering
\includegraphics{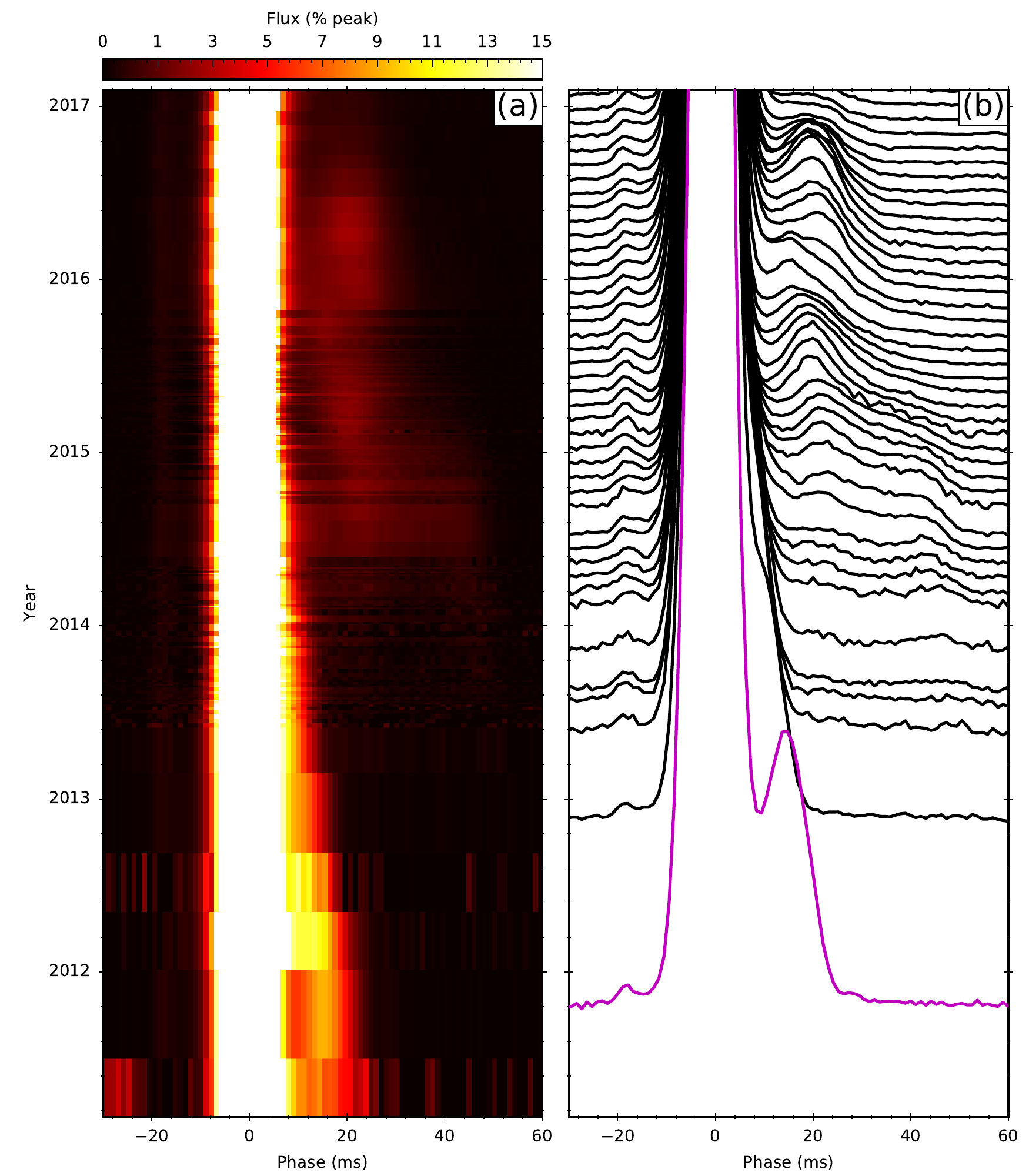}
\caption{Profile evolution of PSR~B2217+47 at $150$\,MHz during the 6-year LOFAR observing campaign. 
All the profiles are normalised to a peak flux of unity and aligned by cross-correlating with a standard template.
LOFAR was in a test phase until 2013 and observed with lower cadence and sensitivity. 
In Panel (a), the profiles are saturated to 15 per cent of the peak intensity.
Panel (b) highlights the complex profile evolution, with profiles averaged every 30 days and only those with $\text{S/N}>1000$ included.
The magenta profile is from \citet{Pil16}.
}
\label{fig:LOFAR_profiles}
\end{figure*}

\subsection{Pulse profile evolution}\label{sec:pulse_ev}
The evolution of the pulse profile in LOFAR observations is shown in Fig.~\ref{fig:LOFAR_profiles}.  
It is very complex, with different components evolving over months, but all {\it trailing} the main peak.
The profile evolution can be divided into two different parts, where the transient component has different characteristics.
\begin{enumerate}
\item\label{bright_component} In early observations, a relatively bright component is present. It shifts towards the main peak until they overlap, reaching the closest approach in 2015. 
\item\label{weak_component} A weaker component with a more complex structure brightens in 2014 and evolves with different characteristics, as highlighted in Fig.~\ref{fig:LOFAR_profiles}b.  
\end{enumerate}
The low S/N and complex profile do not permit a robust model of the weaker component~\ref{weak_component}.
Therefore, the component~\ref{bright_component} was used in most of the analyses and we refer to it as the `bright component'.
None of these transient components is detectable in pulse profiles obtained at $1.5$\,GHz during the same period using the Lovell telescope (Fig.~\ref{fig:Lovell}b), even when using the sensitive GPR method described by \citet{Bro16}.

Furthermore, we detected a weak precursor in LOFAR observations, visible in Fig.~\ref{fig:LOFAR_profiles}b, that has not previously been reported in the literature.  
It is present during the whole observational campaign and does not seem to evolve in time. Therefore, it is likely intrinsic to the (low-frequency) pulsar profile but too weak for previous detection.

\subsubsection{Frequency structure}
In order to study the relative DM value of the transient component and its spectral index, we analysed its evolution as a function of observing frequency.
The observation taken by \citet{Pil16} has been used because in this
early observation the trailing component is relatively bright and well-separated from the
main peak compared to the situation in later observations.  We
manually defined two regions in the pulse profile for the transient component and the main peak, respectively.  

The DM values of the two profile components have been analysed independently.  Their
partial overlap did not permit robust and independent
cross-correlations with standard templates in each
frequency channel.  For this reason, we considered the peak position
of the two components in each frequency channel.  A total of $2948$
frequency channels contained both components brighter than 3 times the noise level.  Considering an uncertainty of one phase bin, a fit to
the frequency structure of the two components gave a statistically insignificant DM difference $\Delta $DM$=(3.9\pm7.0) \times 10^{-3}$\,pc\,cm$^{-3}$.

In order to obtain the spectral index of the transient component, we first calculated its spectral index relative to the main peak.  
The integrated flux of the two profile components was calculated for each frequency channel.  
We then considered the ratio between them so as to remove frequency structures due to the telescope response function, scintillation, RFI, etc.
A fit of these values in different frequency channels gave a relative spectral index of the transient component with respect to the main peak of $\alpha_\text{rel}=-1.60\pm0.03$. \citet{Bil16} used LOFAR measurements together with multi-frequency values from the literature to obtain a spectral index for the main component of PSR~B2217+47 of $\alpha_\text{main}=-1.98\pm0.09$.
Therefore, assuming no spectral turnover, the spectral index of the trailing component can be found using $\alpha_\text{transient}=\alpha_\text{main}+\alpha_\text{rel}=-3.58\pm0.09$.  

The integrated flux of the transient component is 12 per cent that of the main peak at $150$\,MHz.  
Assuming constant values for the obtained spectral indices, these imply a ratio between the two components of $0.3$ per cent at $1520$\,MHz.
Since the pulsar is detected with $\text{S/N}\sim100$ in typical Lovell observations, this steep spectrum explains the lack of detection of the transient component with Lovell.

We repeated the calculation in order to obtain the spectral index of the weaker transient component using 15 high-quality observations where it was relatively well separated from
the main peak.
These observations were recorded between 2014 October 4 and 2016 May 5.  
The average value of the spectral indices is $\alpha_\text{transient}=-3.73\pm0.06$, with all the 15 single values within two sigma from the average. 
This value is compatible with the spectral index of the bright transient component.

\subsubsection{Single-pulse analysis}\label{sec:sp}

\begin{figure}
\centering
\includegraphics{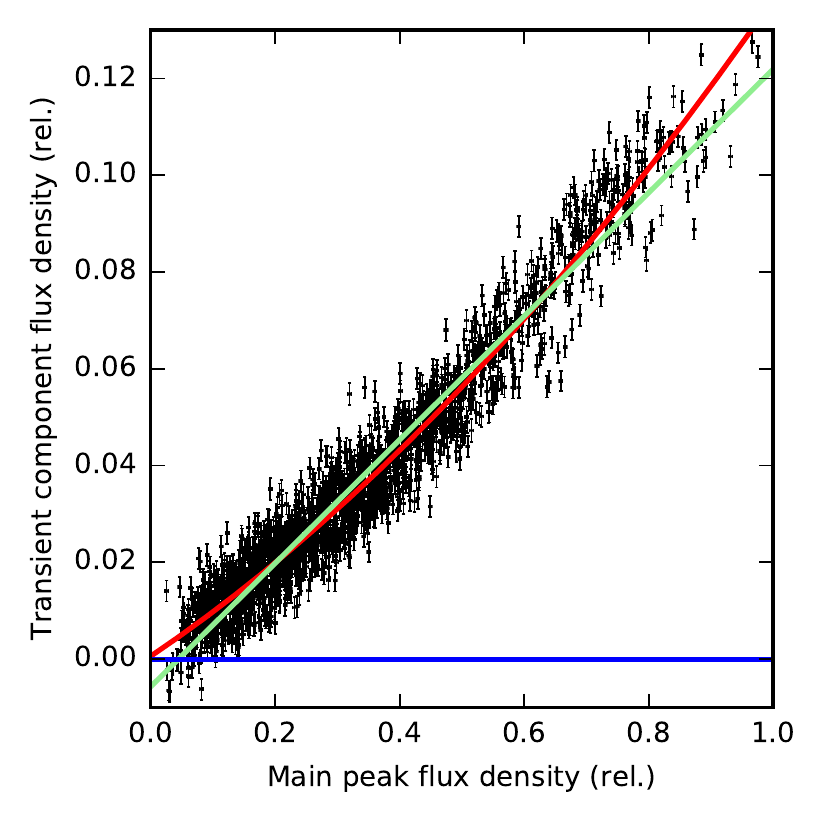}
\caption{Scatter plot of the flux density of the transient component and main pulse peak in each single pulse of a 15-min LOFAR observation.
Units are relative to the brightest pulse peak. 
The green and red lines are linear and quadratic fits to the points, respectively.
The horizontal blue line represents the null flux density for the transient component and the points below are due to noise oscillations.
}
\label{fig:sp}
\end{figure}

We studied the properties of single pulses from PSR~B2217+47 using the observation reported by \citet{Pil16}.
It includes 1769 single pulses from the pulsar.
We found a strong correlation between the flux density of the main peak and of the bright transient component in the individual single pulses detected in the observation (Fig.~\ref{fig:sp}), with a Pearson correlation coefficient of $0.97$.
In order to test this result, we tried the same analysis on components belonging to adjacent single pulses, i.e. correlating the flux density of the transient component in each pulse with the flux density of the main peak in the next pulse.
The correlation disappeared in this additional analysis.
The slight deviation from a linear fit of the points visible in Fig.~\ref{fig:sp} may indicate a second-order effect or could be instrumental in origin.

\subsubsection{Polarisation measurements}

\begin{figure}
\includegraphics{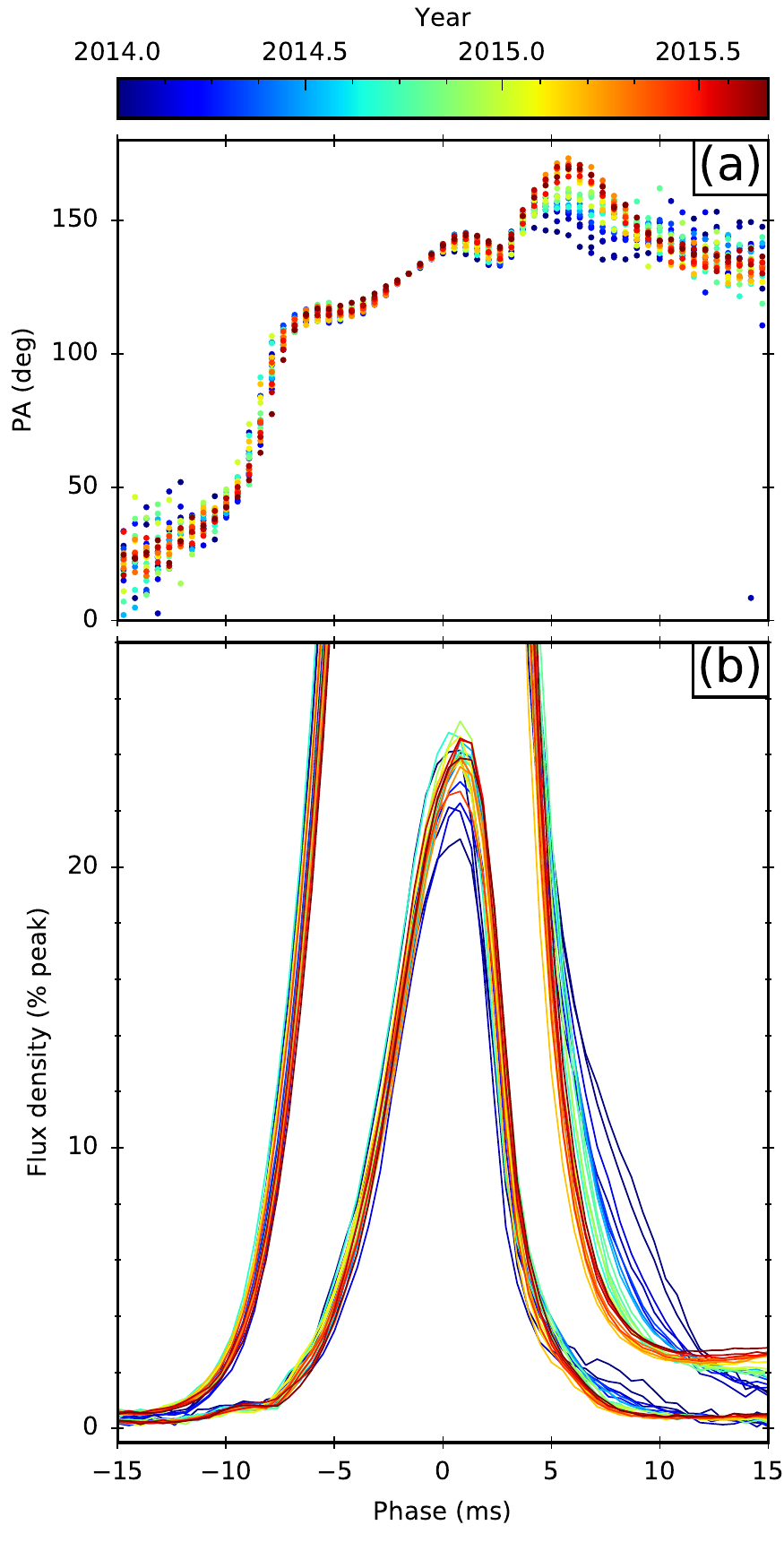}
\caption{Evolution of the polarisation properties of PSR~B2217+47.
PA curves are shown in Panel (a).
Pulse profiles of total intensity (brighter profiles) and linear polarisation (weaker profiles) are shown in Panel (b).
The profiles have been normalized to the peak value of the total intensity for each observation.
The colour scale indicates the epoch of the observations.
}
\label{fig:PA}
\end{figure}

The same LOFAR observations used to calculate flux densities in \textsection\ref{sec:results_DM} were used to study polarisation properties.
One of the observations was taken at an earlier epoch and was excluded from the plots for clarity.
Fig.~\ref{fig:PA}a shows the polarisation angle (PA) curves.
Since a reference polarised signal was not used, an absolute polarisation calibration was not possible to achieve.
Therefore, PA curves were rotated to align at an arbitrary point in the middle of the plot in order to study the relative slope. 
Fig.~\ref{fig:PA}b reports the pulse profiles obtained from the same observations for total intensity and linear polarisation.
The obtained profiles are compatible with those reported in \citet{Nou15}.

A random scatter of the values can be observed along the PA curves in Fig.~\ref{fig:PA}a, except for a specific area in the trailing edge of the main peak, which approximately corresponds to the position of the brighter transient component.
A higher flux of the transient component corresponds to a lower PA value in this region.
In addition, the linear polarisation fraction (L/I) across the profile does not evolve significantly in time. 
The leading part of the profile shows $\text{L/I} \sim 50\%$, while in the rest of the profile $\text{L/I} \sim 20\%$.

\section{Discussion}\label{sec:discussion}


It is possible that cases of pulse profile evolution are fairly common among pulsars but are subtle and therefore difficult to observe.
Pulse profile evolution could be magnified at lower frequencies where it may be detected
in bright pulsars and with sensitive telescopes, as in our case.  
This hypothesis can be tested by regular monitoring of bright pulsars and will be further verified by the next generation of sensitive radio telescopes, in particular the Square Kilometre Array \citep[SKA,][]{Han15}.

Here we consider three effects to explain the observed long-term evolution
of PSR~B2217+47's average pulse profile: (i) a change in the viewing geometry due to pulsar
precession \citep[as considered by][]{Sul94}, (ii) an intrinsic
variation in pulsar emission related to changes in the magnetosphere \citep[similar to those found for a sample of pulsars by][]{Lyn10} or
(iii) variations due to intervening structures in the IISM \citep[perhaps similar to the pulse echoes
seen in the Crab pulsar, and associated with filaments in its surrounding nebula, by][]{Bac00,Lyn01}.

\subsection{Pulsar precession}\label{sec:discussion_precession}


A smooth profile evolution is expected from precession due to the
variation of viewing angle towards the magnetic axis and pulsar beam \citep{Cor93}.
Also, two different effects contribute to timing noise: the additional pulsar spin induced by precession and the fluctuation of the torque due to the change in the angle between the spin and magnetic axes \citep{Cor93}.
In principle, pulsar precession could be a good explanation both for
the structure of the timing residuals (timing noise) reported in
Fig.~\ref{fig:timing}a and for the smooth evolution of the average
pulse profile shown in Fig.~\ref{fig:LOFAR_profiles}.  However, the two effects happen on very different timescales and thus they are difficult to attribute to the same precession period, even if non-axisymmetric precession of the neutron star is present.
Furthermore, the precession model does not explain the strong DM variations we detect, which would then be coincidental.
Given the paucity in the literature of DM variations having the measured intensity in this short period, such a coincidence seems unlikely.  
In addition, all the observed profile variations only affect the trailing part of the profile, while the shape and relative position of the weak precursor and the main peak are steady.
This would be an additional coincidence in this scenario.
Moreover, this model does not explain the correlation between the flux density of the main peak and of the bright transient component found in \textsection\ref{sec:sp}.

\subsection{Variations in pulsar emission}\label{sec:discussion_emission}

Perturbations in the plasma filling the magnetosphere can cause a variation both in the average pulse profile (due to changes in particle flux and currents) and in the spin-down evolution of the neutron star \citep[due to changes in the electromagnetic torque,][]{Spi04,Kra06,Lyn10}.
It is unclear whether the spin-down variations that we measure for PSR~B2217+47 and reported in Fig.~\ref{fig:timing}b are related with the observed profile variations visible in Fig.~\ref{fig:LOFAR_profiles}. 
However, a model with two or three separate $\dot{\nu}$ levels as proposed by \citet{Lyn10} is clearly inconsistent with our results. 

Three models possibly related to changes in the pulsar's magnetosphere were considered, namely the nested cone model \citep{Ran83}, the refractive model \citep{Bar86} and an asteroid encounter \citep{Cor08}, but none of these models readily explain the observed asymmetry in the pulse profile evolution, the strong DM variations detected, or the correlation found between the flux density of the main peak and of the transient component.

\subsubsection{The role of the glitch}
A connection between pulse profile variations and spin-down evolution
has been observed by \citet{Wel11} in PSR~J1119$-$6127, where a
temporary change in the pulse profile was contemporaneous with, and
thus likely induced by, a glitch in the pulsar's rotation.
The connection is interesting to consider in light of the glitch detected in close proximity to the \citet{Pil16} observation.
However, the profile of PSR~J1119$-$6127 changed on a weekly time-scale as opposed to the yearly variations observed in PSR~B2217+47.
An analysis of the pulse profile evolution in PSR~B2217+47 shows that the occurrence of the glitch only coincides with a large separation between the bright transient component and the main pulse, but transient profile components were also detectable before the glitch occurred. 

\citet{Wel11} also detected a few strong pulses from PSR~J1119$-$6127, at a particular pulse phase where no components were present in the integrated pulse profile, in the first single-pulse resolved observation after the glitch.  
We performed a search for bright single pulses away from the profile components in the LOFAR observation presented by \citet{Pil16}.  
We did not find any pulse above 6 per cent of peak intensity in that observation before a phase of $-0.05$ or after a phase of $0.04$ with respect the main peak.

\subsection{Structures in the IISM}
A structure in the IISM close to the LoS could create an additional transient component in the pulse profile by deflecting radio waves back to the observer.
The leading edge of the profile is expected to be unaffected by IISM propagation, as is observed in all the profile variations detected in PSR~B2217+47 \citep[including the one reported by][]{Sul94}. 
This is a firm prediction to test this hypothesis with future observations.

The transient component will have a delay with respect to the direct signal and its evolution with time is analysed in Appendix~\ref{sec:appendix}.
In most of the practical cases, this evolution will be quadratic in time (Eq.~\ref{eq:delay_end}).
Despite the fact that only the bright transient component could be analysed in detail, it is plausible to assume that the multiple transient components observed have the same origin.
In the IISM interpretation, they could arise from a group of structures or from inhomogeneities in a large structure passing close to the LoS.
Since the scattering process is highly chromatic, the steep spectrum of the transient components is also expected within this model.
The strong DM variations presented in Fig.~\ref{fig:timing}c and \ref{fig:LOFAR_DM_flux}a prove that there is significant variability in the IISM towards this pulsar and the same IISM structures can explain both the effects.
A strong support to this model is given by the correlation in single pulses between the flux density of the bright transient component and the main peak discussed in \textsection\ref{sec:sp}.
This is also expected if the transient component is an echo of the main peak.

\begin{figure}
\includegraphics[width=\columnwidth,height=.8\textheight,keepaspectratio]{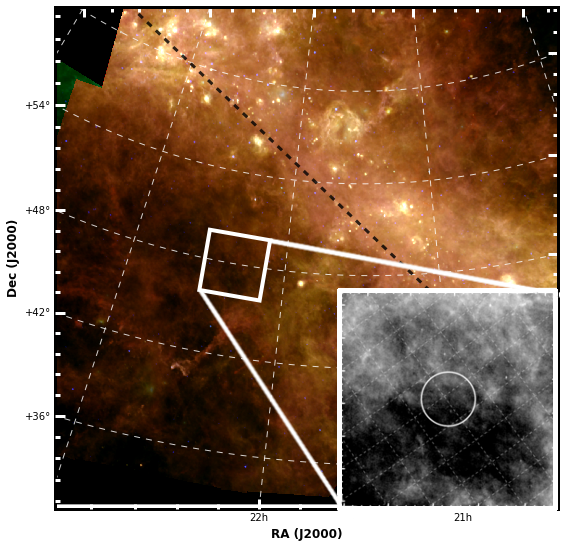}
\caption{A three colour mosaic of images of the field around PSR~B2217+47 \citep{Miv06} obtained with the \textit{Infrared Astronomical Satellite (IRAS)} through the Improved Reprocessing of the IRAS Survey (IRIS).
Red is the $100$\,$\mu$m band, green $60$\,$\mu$m and blue $25$\,$\mu$m.  
The dashed black line indicates the Galactic Plane.  
The inset is a $60$\,$\mu$m \textit{AKARI} observation \citep{Mur07}. 
The white circle is centred on the pulsar location and has a radius of $30\arcsec$.  
The white box and \textit{AKARI} image are $4\times4$\,arcmin$^2$.}
\label{fig:IISM_image}
\end{figure}

Fig.~\ref{fig:IISM_image} shows an infrared image of the field of PSR~B2217+47.
Clouds of dust and gas are seen to extend from the Galactic plane.
This is thus an inhomogeneous region of the neutral interstellar medium where ionised structures could possibly form.
On the contrary, there is no clear indication of a bright, compact structure near the pulsar.
Also, no obvious structure is visible in the H$\alpha$ map presented by \citet{Fin03}.
X-ray images of the field obtained by \textit{ROSAT} \citep{Vog99} do not show any significant emission from the pulsar region either.

\begin{figure}
\includegraphics{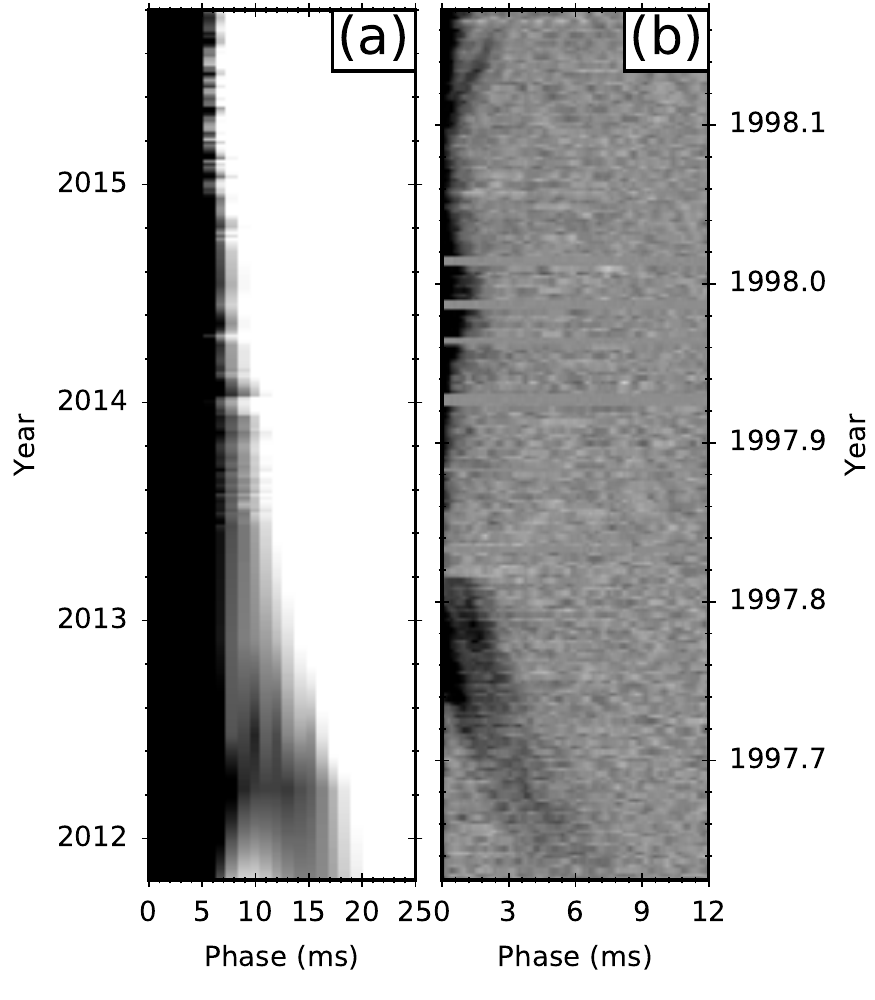}
\caption{Comparison between the profile evolution of PSR~B2217+47 (a) and of the Crab pulsar \citep[b,][]{Lyn01}.
The profiles in Panel (a) are clipped between 5\% and 15\% of the peak intensity, interpolated with a linear spline and plotted with a logarithmic colour scale to highlight the evolution of the brightest transient component.
}
\label{fig:echo}
\end{figure}

An echo in a pulsar's pulse profile has previously been observed in PSR~B0531+21 \citep{Bac00,Lyn01}, although in that case the structures are believed to be within the Crab nebula itself and not the IISM.
Fig.~\ref{fig:echo} shows a visual comparison between the evolution of
the bright transient component in PSR~B2217+47 and a Crab echo
event reported by \citet{Lyn01}.  
The evolution of the two transient components presents some similarities.  
The feature in the Crab shows a parabolic shape in time versus rotational phase (as expected from Eq.~\ref{eq:delay_end}) and the separation of the component in PSR~B2217+47 seems consistent with such an approach, although the S/N and number of observations in the earlier eras is insufficient to precisely quantify this.
Also, the relative fractional power in the transient components is of the same order in the two cases, between $\sim 5$\% and 10\% of the integrated pulse flux.  
The different timescales of the two evolutionary paths can be ascribed to different relative distances between the structure, the pulsar and Earth (Eq.~\ref{eq:delay_end}).
The absence of a symmetric receding arm in the evolution of the bright transient component of PSR~B2217+47 may be due to a non-spherical geometry of the IISM structure (e.g. elongated in one or two directions). 
About 30\% of the Crab echoes reported by \citet{Lyn01} only show one of the parabolic arms.  
It is interesting to note that Os{\l}owski et al. (in prep.) observe with LOFAR a low-frequency profile evolution of PSR~B1508+55 that presents various similarities to the one reported here.
Three transient components appear at different epochs and shift quadratically in phase over years with the original detection corresponding to a much larger delay than presented here.
They also attribute the evolution to IISM effects. 

The IISM model cannot entirely account for the strong timing noise observed in PSR~B2217+47.
Despite the fact that small profile variations can cause timing noise \citep[e.g.][]{Len17},
most of the timing observations used here were acquired around $1.5$\,GHz, where the flux inferred for the transient component is less than $0.3$\% of the main pulse, and a GPR analysis found no evidence for profile variability (\textsection\ref{sec:pulse_ev}).
In addition, the peak-to-peak variations in the timing residuals is $\sim 30$\,ms compared to a main component width of $\sim 8$\,ms at 1.5\,GHz.  
Furthermore, the presence of timing noise in many slow-spinning pulsars \citep{Cor80,Hob04} supports the hypothesis that the two effects are unrelated.

\section{Modelling the IISM structure}\label{sec:IISM}
Among the three interpretations we considered in the previous section, the IISM model is the one that best explains the characteristics of the observed profile evolution of PSR~B2217+47.
Here, we use the observations to infer some properties of the IISM structure possibly causing the profile evolution.
We calculate a distance to the structure from Earth by using Eq.~\ref{eq:distance} and the separation in pulse phase of the bright transient component (Fig.~\ref{fig:LOFAR_profiles}).
Around $t=3.5\pm0.3$~years before the closest approach, the delay of the transient component is
$\tau=14.5\pm0.5$\,ms.  
The overlap with the main peak suggests that the IISM structure nearly crosses the LoS, with a delay $\tau_*=1\pm1$\,ms.  
The timing solution presented in Table~\ref{tab:ephemeris} implies a pulsar proper motion $\mu=20.6\pm2.9$\,mas\,yr$^{-1}$ and the pulsar distance is estimated to be $r=2.2\pm0.3$\,kpc using the \textsc{NE2001} model \citep{Cor02}. 
The proper motion of the IISM structure is assumed to be negligible since the pulsar speed is estimated to be much larger than typical IISM velocities \citep[e.g.][]{Hil05} and the Earth's orbital speed.
From Eq.~\ref{eq:distance}, the resulting distance of the IISM structure from Earth is $d=1.1\pm0.2$\,kpc, approximately half the pulsar distance. 
The error is dominated by the uncertainty on the pulsar's proper motion.
The distance of the structure implies a deflection angle $\delta=150\pm25$\,mas.

We estimated the transverse radius of the IISM structure that causes the DM variation detected by LOFAR:
\begin{equation}\label{eq:radius_structure}
R\approx d\mu t,
\end{equation}
where $t$ in this case is the time span during which the variation is present.
Fig.~\ref{fig:LOFAR_DM_flux}a shows the density profile of the IISM causing the DM variation along the axis parallel to the pulsar's transverse velocity.  
The exact structures causing the DM and profile variation may not be the same, but are likely to belong to the same group, i.e. to be close in space with respect to the relative distances of the pulsar, IISM structure and Earth.
Therefore, we can substitute the distance from the IISM structure calculated above.  
The largest DM variation in LOFAR data lasts for $t\gtrsim 1.5$\,yr, which implies $R\gtrsim 34$\,AU.  Under the assumption that the system scale does not change significantly in two decades, we can
repeat the calculation for the DM variation reported by \citet{Ahu05}.
Considering the bulk of the DM variation to last for $t\approx400$
days, we obtain $R\approx25$\,AU.
The observed DM variations imply that most of the IISM structures are
over-dense, in agreement with \citet{Rom87} and \citet{Ban16}, and in contrast to \citet{Pen12}, although one
under-dense region is apparent in Fig.\ref{fig:timing}c around MJD 51000, right before the DM variation reported by \citet{Ahu05}.

\subsection{Ionised blobs}\label{sec:blobs}
We consider the possibility that the observed profile variations are due to deflection by approximately axially-symmetric IISM structures \citep{Wal98}.
We calculate the implied electron density using both DM and profile variations.
From the DM variation:
\begin{equation}\label{eq:dDM}
\Delta\text{DM}=L\overline{n}_e,
\end{equation}
where $\overline{n}_e$ is the average electron density of the IISM structure and where $L$ is the radius of the structure along the LoS.
If we assume spherical symmetry, $L\approx R$ and the relation can be combined with Eq.~\ref{eq:radius_structure} to obtain:
\begin{equation}
\overline{n}_e \approx \frac{\Delta\text{DM}}{d\mu t}.
\end{equation}
From Fig.~\ref{fig:LOFAR_DM_flux}a, we estimate $\Delta\text{DM}\approx4 \times 10^{-3}$\,pc\,cm$^{-3}$, which implies $\overline{n}_e\approx25$\,cm$^{-3}$.
The same calculation for the DM variation reported by \citet{Ahu05}, $\Delta\text{DM}\approx0.02$\,pc\,cm$^{-3}$, implies $\overline{n}_e\approx170$\,cm$^{-3}$.

Considering a refractive plasma lens, \citet{Hil05} found the relation
between the mean electron density of the lens $\overline{n}_e$ and the
refracting angle $\delta$:
\begin{equation}
\overline{n}_e = \frac{5.4\delta}{\lambda^2}\,\text{m}^2\,\text{mas}^{-1}\,\text{cm}^{-3},
\end{equation}
where $\lambda$ is the observing wavelength, which varies between $\sim 1.5$~--~$2.5$\,m across the LOFAR HBA band.
Substituting the values we calculated in the previous section, we find $\overline{n}_e\sim130$~--~$360$\,cm$^{-3}$ across the band.
This value is in good agreement with the mean electron density inferred from the DM variation reported by \citet{Ahu05}.  
The obtained electron densities are roughly consistent with standard ESE models \citep[e.g.][]{Mai03,Hil05}.

\subsection{IISM structures around hot stars}
During the writing of this manuscript, \citet{Wal17} presented evidence for ionised clouds around hot stars causing intra-day variability of radio quasar fluxes.
If confirmed, the same structures could generate ESEs and explain the high electron density required.
The distance estimated above for the IISM structure causing the profile variation of PSR~B2217+47 is $\sim1$\,kpc, under the assumption that the pulsar proper motion is much higher than the proper motion of the structure.
However, structures connected with a star will have approximately its same velocity, usually larger than the average speed of the interstellar medium \citep[e.g.][]{Lee07}.
Unfortunately, the large uncertainty in the distance estimated from Eq.~\ref{eq:distance} does not permit to put constraints on the expected star's distance.

Following \citet{Wal17}, we searched for hot stars (spectral types O, B and A) closer than 2\,pc to PSR~B2217+47's LoS in the \textit{Hipparcos} catalogue \citep{Per97}.
Unfortunately, stellar spectra are not available at distances $\gg100$\,pc.
Only stars with a reliable parallax measurement (i.e. larger than the 2-$\sigma$ value) were selected.
Three stars met these criteria, all with a spectral type A.
\begin{enumerate}
\item HIP~110422 is at $101\pm6$\,pc and $\sim1.9$\,pc from the LoS.
\item HIP~110253 is at $230\pm30$\,pc and $\sim1.8$\,pc from the LoS.
\item HIP~110139 is at $350\pm70$\,pc and $\sim1.8$\,pc from the LoS.
\end{enumerate}
In order to evaluate the probability of coincident alignment, we considered a $15\times15$\,deg$^2$ sky region centred on the pulsar position.
A total of 387 hot stars were present in the catalogue in this area.
Selecting $10^6$ random positions in this area, we found that on average $\sim1$ hot star was closer than 2\,pc to the LoS, in agreement with \citet{Wal17}.
We conclude that this is an interesting hypothesis but our data does not permit to constrain it. 

\subsection{Scintillation arcs and inclined sheet model}\label{sec:sheets}
We considered the possibility that the observed profile variation is an effect similar to scintillation arcs.
Similar IISM structures could in principle account for both effects.
In fact, scintillation arcs could be caused by structures closer to the LoS that deflect higher-frequency radio waves at lower deflection angles. 
The delay of the deflected waves will be shorter and they will interfere with the direct waves.
The same structures might deflect lower-frequency waves at higher deflection angles when farther away from the LoS.
In this case, the delay of the deflected waves will be longer and they will create additional components in the pulse profile, trailing the main peak.

Compared to blobs, corrugated plasma sheets that are nearly aligned with the LoS require a lower density to deflect radio waves at the required angle \citep{Pen14}.  
The same argument can be used in the current study to explain the strong DM variations that were detected.  
\citet{Sim17} described a typical plasma density in the sheet on the order of $n_e=0.3$\,cm$^{-3}$.  
Substituting into Eq.~\ref{eq:dDM}, this implies a length $L=10^4$\,AU from the DM variation reported by \citet{Ahu05} and $L=3\times10^3$\,AU from the more recent DM variation measured with LOFAR.
The radius $\text{R}\sim30$\,AU calculated in the previous section would represent, in this scenario, the projected width of the sheets.
It is unclear whether these values are in agreement with inclined sheet models.

\section{Conclusions}\label{sec:conclusions}
We presented the analysis of pulse profile evolution of PSR~B2217+47, similar to an event previously reported by \citet{Sul94}. 
We performed an intensive, multi-year observing campaign with LOFAR ($\sim150$\,MHz) and Lovell  ($\sim1.5$\,GHz) telescopes.
A smooth but complex evolution of the pulse profile was observed during the entire duration of the LOFAR observing campaign.
In addition, we report for the first time the presence of a weak precursor visible in LOFAR
observations, which remains stable for the whole range of our observing campaign.
No profile evolution was detected during the same timespan at $1.5$\,GHz with the Lovell telescope.
Significant variations in the ionised interstellar electron density towards this pulsar are possibly related and are discussed in a companion paper (Donner et al., in prep.).
An earlier episode of large interstellar density variations was reported by \citet{Ahu05}.

Both pulsar precession, considered as a possible explanation for the observed profile evolution by \citet{Sul94}, and an intrinsic variation in the pulsar emission do not explain the coincident DM variations, the stability of the leading edge of the profile, and a correlation between the flux density of the transient component and of the main peak in the single pulses of one observation.
On the other hand, these characteristics can be attributed to propagation effects in the IISM.
Also, there is evidence for a quadratic approach of the transient component to the main peak, which is expected in this model, although the quality and cadence of early observations do not permit a robust analysis.
Future high-resolution images at $150$\,MHz could spatially resolve the IISM structure, and thereby confirm this interpretation.
We inferred some of the properties of the putative IISM structures causing DM and profile variations and discussed the two main models to explain ESEs \citep{Wal98,Pen12} in this context in \textsection\ref{sec:blobs} and \textsection\ref{sec:sheets}.

Time-variable dispersion and profile-shape changes affect both the precision and accuracy of pulsar timing experiments \citep[e.g.][]{Cog93}. 
IISM variations far less significant than those observed in PSR~B2217+47 are already significantly affecting most of the currently existing high-precision pulsar-timing data sets \citep{Ver16}.
It is therefore a timely effort to investigate and characterise these events and to constrain their origins and occurrence rates, in order to allow their proper modelling and treatment in the decades to come.

\section*{Acknowledgements}
We thank Joanna Rankin, Thomas Scragg and Joel Weisberg for helpful discussions, Vlad Kondratiev for assistance processing LOFAR observations and James McKee for reviewing the manuscript before submission.
DM and JWTH acknowledge funding from the European
Research Council under the European Union's Seventh Framework
Programme (FP/2007-2013) / ERC Starting Grant agreement nr. 337062
(`DRAGNET').  JWTH also acknowledges funding from an NWO Vidi
fellowship.
This paper is based in part on data obtained with the International LOFAR Telescope
(ILT). LOFAR (van Haarlem et al. 2013) is the Low Frequency Array
designed and constructed by ASTRON. It has facilities in several
countries. These are owned by various parties (each with their own
funding sources) and are collectively operated by the ILT
foundation under a joint scientific policy.
This paper made use of data from the Effelsberg (DE601) LOFAR station
funded by the Max-Planck-Gesellschaft; the Tautenburg
(DE603) LOFAR station funded by the BMBF Verbundforschung project
D-LOFAR I and the European Union (EFRE); and the J\"ulich (DE605)
LOFAR station supported by the BMBF
Verbundforschung project D-LOFAR I. The
observations of the German LOFAR stations were carried out in the
stand-alone GLOW mode (German LOng-Wavelength array), which is
technically operated and supported by
the Max-Planck-Institut f\"ur Radioastronomie, the Forschungszentrum
J\"ulich and Bielefeld University.
LOFAR Station FR606 is hosted by the Nan\c{c}ay Radio Observatory, which
is operated by Paris Observatory, associated with the French Centre
National de la Recherche Scientifique (CNRS) and Universit\'e d'Orl\'eans.
The Rawlings Array is operated by LOFAR-UK as part of the International LOFAR Telescope, and is funded by LOFAR-UK and STFC. 
Pulsar research at Jodrell Bank and access to the Lovell Telescope is supported by a Consolidated Grant from the UK's Science and Technology Facilities Council.
The Long Baseline Observatory is a facility of the
National Science Foundation operated under cooperative agreement by
Associated Universities, Inc. 

\bibliographystyle{mnras}
\bibliography{michilli}

\appendix
\section{Evolution of the time delay of an echo in the pulse profile}\label{sec:appendix}
\begin{figure}
\includegraphics[width=\columnwidth,height=.8\textheight,keepaspectratio]{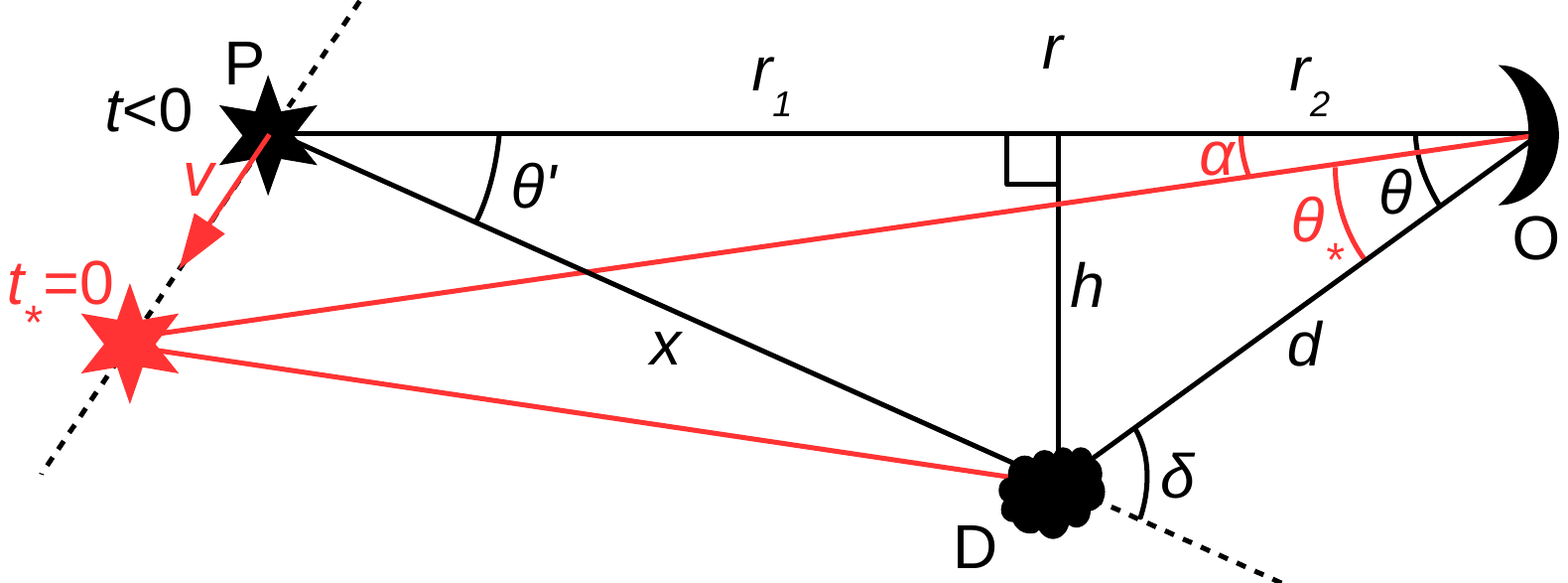}
\caption{Schematic representation of the geometry causing an echo in the pulse profile (not to scale).  
A density variation in the IISM (D) deflects at an angle $\delta$ radio waves emitted from a pulsar (P) back to the observer (O).
The pulsar is moving with a velocity $v$ with respect to the IISM structure and thus the angles $\alpha$, $\theta$ and $\theta'$ evolve with time. 
Two different epochs are represented: red symbols indicate the system configuration at the closest approach, i.e. when $\theta$ and $\theta'$ are smallest, while black symbols refer to an arbitrary time $t$ ($t<0$ in this example).
The distance $h$ of the LoS to the IISM structure divide the distance $r$ from Earth to the pulsar in two segments $r_1$ and $r_2$. $x$ is the distance between the pulsar and the IISM structure D.}
\label{fig:geom_delay}
\end{figure}

When a density change in the IISM deflects radio waves from a pulsar
back to the LoS, an echo in the pulse profile may be
detected.  The delay between the echo and the normal components in the
pulse profile is due to the additional path length.  This is depicted in
Fig.~\ref{fig:geom_delay}, where a pulsar is moving in the barycentric reference system with respect to a structure in the IISM. 
While a blob of plasma is represented in the cartoon, this model accounts for any deflection that a structure in the IISM may produce.
In fact, the discussion is based on geometrical considerations and
does not depend on the proprieties of the IISM structure or of the
deflecting mechanism.  We calculate the distance $d$ between the
observer and the IISM structure by considering two values of the echo delay at
different times.  These are produced by different geometrical
configurations of the system indicated with black and red colours in
the illustration.

In Fig.~\ref{fig:geom_delay}, the geometrical delay between direct and deflected radio waves is given by \begin{equation} \tau = \frac{d+x-r}{c}, \label{eq:delay_def}\end{equation}
where $c$ represents the speed of light and the other quantities are defined in Fig.~\ref{fig:geom_delay}.
Noting that $r=r_1+r_2$, this can be rewritten as
\begin{equation} \tau = \frac{x}{c} \left(1-\cos{\theta'}\right) + \frac{d}{c} \left(1-\cos{\theta}\right). \end{equation}
In all practical cases, the angles $\theta$ and $\theta'$ are very small, therefore we can approximate the relation as
\begin{equation} \tau \approx \frac{x\theta'^2+d\theta^2}{2c}. \end{equation}
In the small-angle approximation, the sine law gives $x\theta' = d\theta$ which, substituted in the previous equation, yields
\begin{equation} \tau \approx \frac{\theta^2d}{2c}\left(\frac{d}{x}+1\right). \label{eq:delay_end}\end{equation}
In equation~\ref{eq:delay_end}, the only quantity changing in the considered timespan is $\theta$. 
We can assume that $\theta$ evolves at a nearly constant rate given the short path travelled by the pulsar compared to the system scale.
This implies a quadratic variation of the delay in time across the pulse profile.

To find the distance $d$ as a function of $\tau$ and $\theta$, we can substitute $x$ with its expression from the cosine law, after applying the small-angle approximations.
This, however, leads to a cubic equation in $d$ difficult to invert.
We calculate approximated equations considering three different configurations: (i) an IISM structure local to the pulsar, (ii) near half-way between the pulsar and the Earth and (iii) local to the Earth.
The delay can be expressed as
\begin{equation}\label{eq:tau_approx}
\begin{array}{llll}
d\approx r &\Rightarrow& \tau\approx \frac{r\theta^2}{2c}\left(\frac{r}{x}+1\right)& \text{local to pulsar} \\
\theta\approx\theta' &\Rightarrow& \tau\approx \frac{dr\theta^2}{2c(r-d)}& \text{half-way} \\
x\approx r &\Rightarrow& \tau\approx \frac{d\theta^2}{2c}\left(\frac{d}{r}+1\right)& \text{local to Earth,}
\end{array}
\end{equation}
where we made use of the relations $r_1=x\cos{\theta'}$ and $r_2=d\cos{\theta}$ and ignored terms of higher order to obtain the relation for the half-way configuration.

The angle $\theta$ can be estimated by measuring delay values at two different epochs if one corresponds to the closest approach.
We denote quantities measured at the closest approach with an asterisk.
The angle $\alpha$ can be expressed as $\alpha=\mu t$, where $\mu$ is the pulsar proper motion and $t$ is the time between the two observation ($t_*=0$).
Since the triangle formed by the pulsar at the two epochs and the structure is right-angled, we obtain $\theta^2=\theta_*^2+\mu^2t^2$.
Evaluating Eq.~\ref{eq:delay_end} for $\tau$ and $\tau_*$ we get
\begin{equation}\label{eq:theta}
\theta=\mu t\sqrt{\frac{\tau}{\tau-\tau_*}}.
\end{equation}
For structures that cross the LoS, $\tau_* = 0$ and hence Eq.~\ref{eq:theta} reduces to $\theta = \mu t$.

By inverting Eqs.~\ref{eq:tau_approx} and substituting Eq.~\ref{eq:theta}, we obtain
\begin{equation}\label{eq:distance}
\begin{array}{ll}
d \approx r-\frac{r^2\mu^2t^2}{2c(\tau-\tau_*)-r\mu^2t^2}& \text{local to pulsar} \\
d \approx \frac{2cr(\tau-\tau_*)}{2c(\tau-\tau_*)+r\mu^2t^2}& \text{half-way} \\
d \approx \sqrt{\frac{r^2}{4} + \frac{2cr(\tau-\tau_*)}{\mu^2t^2}} - \frac{r}{2}& \text{local to Earth}
\end{array}
\end{equation}
Finally, from geometrical considerations and the assumptions made above,  the deflection angle $\delta$ is given by
\begin{equation}
\begin{array}{ll}
\delta \approx \frac{\pi-\theta}{2} = \frac{\pi}{2}-\frac{\mu t}{2}\sqrt{\frac{\tau}{\tau-\tau_*}} & \text{local to pulsar} \\
\delta \approx 2\theta = 2\mu t\sqrt{\frac{\tau}{\tau-\tau_*}} & \text{half-way} \\
\delta \approx \pi-\theta = \pi-\mu t\sqrt{\frac{\tau}{\tau-\tau_*}}& \text{local to Earth}
\end{array}
\end{equation}

We applied Eqs.~\ref{eq:distance} to the echoes reported by \citet{Lyn01} in the Crab pulsar in order to have a comparison between different models.
Considering a delay of 7\,ms 50 days before the closest approach, \citet{Lyn01} find a distance between the pulsar and the IISM structure of $1.3$\,pc.  
\citet{Gra11} find a distance for the same event of $5$\,pc, although they argue that this result could be overestimated.
Using their same parameters, we find $x\sim2.5$\,pc using a pulsar distance of $1.7$\,kpc derived by \citet{Cor02}.
However, this result does not consider the motion of the lens with respect to the pulsar's LoS.
Given also the large uncertainty expected from the limited knowledge of the pulsar distance and proper motion, we consider the three values to be reasonably consistent.

\bsp	
\label{lastpage}
\end{document}